\documentclass[a4paper]{article}           % Switch to report for front page

\usepackage{amsmath,amsfonts,amsthm,amssymb}
\usepackage[utf8]{inputenc}
\usepackage[T1]{fontenc}
\usepackage{setspace}                      % Allows more fine-grained control over line spacing
\usepackage{fancyhdr}                      % Headers and footers
\usepackage{lastpage}                      % Allows references to the last page of the document
\usepackage{chngpage}                      % Change format mid-page ?
\usepackage{soul}                          % Highlights text, with \hl{}
\usepackage[usenames,dvipsnames]{color}    % Add color to text
\usepackage{graphicx,float,wrapfig}
\usepackage{ifthen}                        % \ifthenelse{}
\usepackage{listings} 
\usepackage{hyperref}
\usepackage[usenames,dvipsnames]{color}
\usepackage{subfig}
\usepackage{placeins}

\newtheorem{mydef}{Definition}
\newcommand{\Title}{The Arduino as a Hardware Random-Number Generator}
\newcommand{\SubTitle}{Final Report}
\newcommand{\DueDate}{December, 2011} % Or \today
\newcommand{\Class}{Ardrand}
\newcommand{\ClassInstructor}{Ýmir Vigfússon}
\newcommand{\AuthorName}{Benedikt Kristinsson}
\newcommand{\DueLang}{}     % Icelandic   (perhaps some ifelse on language pack)
\definecolor{MyDarkGreen}{rgb}{0.0,0.4,0.0}
\definecolor{MyDarkRed}{rgb}{0.4,0.0,0.0}

\newcommand{\tmpsection}[1]{}
\let\tmpsection=\section

\renewcommand{\section}[2]{

    \ifthenelse{
      \equal{#2}{*} % I have to be oddly specific here
    }
    {
      \tmpsection{References}
      \tmpsection{\sc{#2} }
    }
    {\tmpsection{\sc{#1} } }

}

\title{
    \Class:\ \Title
    }
\date{\small{\DueLang\ \DueDate}}
\author{\AuthorName\\Advisor: \ClassInstructor\\Reykjavik University}

\begin{document}
\maketitle

\begin{center}
\textit{For the kid playing space station in the school yard. }
\end{center}

% Uncomment the \tableofcontents and \newpage lines to get a Contents page
% Uncomment the \setcounter line as well if you do NOT want subsections
%       listed in Contents
% Remeber to compile twice
%\setcounter{tocdepth}{1}
%\tableofcontents
%\newpage

%\clearpage
%x\section{Lausn verkefnis og útfærsla}

\begin{abstract}

  Cheap micro-controllers, such as the Arduino or other controllers
  based on the Atmel AVR CPUs are being deployed in a wide variety of
  projects, ranging from sensors networks to robotic submarines. In
  this paper, we investigate the feasibility of using the Arduino as a
  true random number generator (TRNG). The Arduino Reference Manual
  recommends using it to seed a pseudo random number generator (PRNG)
  due to its ability to read random atmospheric noise from its
  analog pins. This is an enticing application since true bits
  of entropy are hard to come by. Unfortunately, we show by
  statistical methods that the atmospheric noise of an Arduino is
  largely predictable in a variety of settings, and is thus a weak
  source of entropy. We explore various methods to extract true
  randomness from the micro-controller and conclude that it should not
  be used to produce randomness from its analog pins.

\end{abstract}

\section{Introduction}
% Motivation

Various aspects in our lives may seem random --- so thinking that generating randomness might seem easy at first glance. But when one inquires further one quickly realizes that due to the deterministic nature of CPUs, it is impossible for them to generate random numbers. 

However, there is a great need for unpredictable values in cryptography. Kerckhoff's principle states that ``a cryptosystem should be secure even if everything about the system, except the key, is public knowledge''. Almost all encryption schemes rely on the notion of secret keys so those keys must be generated in an unpredictable way, or else the encryption scheme is useless. Examples of this are the keystream in a one-time-pad, the primes in the RSA algorithm and the challenges used in a challenge-response system~\cite{menezes1996,anthes2011}. Many secure encryption protocols use nonces (numbers used once) to add ``noise'' to messages ~\cite{anthes2011}. If these numbers are predictable, the nonces do not serve much purpose.

Since regular computers are unable to produce truly random numbers, psuedorandom number generators (denoted PRNG) are the name of the game. A PRNG is a one-way function $f$ the generates random sequnces, of either integers or bits, from an intial seed $s$ and then applies the function iteratively to generate the sequence ~\cite{menezes1996}. In a cryptographic system, a weak source for the seed weakens the whole system. It may allow an adversary to break it, as was perhaps most notably demonstrated by breaking the method that the Netscape browser used to seed its PRNG ~\cite{netscape}. 

Thus a PRNG can only be random if its seed is truly random and its output is only a function of the seed data, the actual entropy of the output can never exceed that of the seed. However, it is generally computationally infeasible to distinguish between a good PRNG and a perfect RNG. A true random number generator (TRNG) uses a non-deterministic source to produce randomness e.g. measuring chaotic systems in nature like thermal noise, shot noise or flicker noise, which are all present in resistors ~\cite{intel}. Using background radiation and a Geiger counter is an appealing option, but expensive\footnote{A simple search on Amazon.com reveals that a USB connected Geiger counter costs about \$300} and thus unavailable for the public. 

\begin{figure}[H]
  \centering  
  \includegraphics[width=0.7\columnwidth]{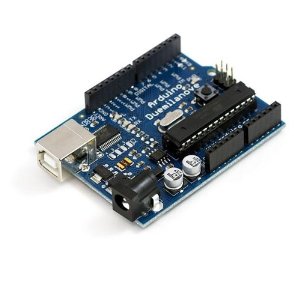}
  
  \caption{Arduino Duemilanove}
  \label{fig:ard3space}
\end{figure}

The Arduino is a free and open-source electronics single-board micro-controller with an Atmel AVR CPU. There are several different versions of the board available\footnote{See: \url{http://arduino.cc/en/Main/Boards}}, but we used the Arduino Duemilanove\footnote{See: \url{http://arduino.cc/en/Main/ArduinoBoardDuemilanove} for full specifications} board (with the ATMega328 ~\cite{atmegads} micro-controller) for this research. The Arduino toolkit has the \texttt{analogRead} function that reads from a given analog pin on the board and returns a 10-bit integer. This function maps input voltages between 5 and 0 volts to integers in the range $[0..1023]$. This is what we tried to use in order to extract entropy.

Micro-controllers like the Arduino are heavily used in e.g. sensor networks ~\cite{tsense} where data integrity is a key issue. It follows that the demand for high quality entropy is rather high in those situations. 

The Arduino Reference Manual suggests that reading from an unconnected analog pin gives a ``fairly random'' number ~\cite{ardref}, ideal for seeding the \texttt{avr-libc} PRNG\footnote{Archival of this claim: \url{http://web.archive.org/web/20110428064453/http://arduino.cc/en/Reference/RandomSeed}}. We will later show that the numbers are generally not random, and that the reading from an unconnected pin provides very limited entropy. We will also show that building a RNG with the Arduino is infeasible and that if you follow the Arduino Reference Manual, the sheer lack of possible seeds makes it relatively easy for an adversary to guess the seed. We will provide a proof of concept tool for doing such guesswork automatically. 

We also attempt to build a random bit generator from the Arduino (without adding extra hardware). We will pose and discuss a few algorithms and discuss how they perform statistically speaking testing and how they fare at extracting entropy. Ultimately, we were unable to identify any such method which rises concerns over the use of Arduino as a TRNG. 

\subsection{Contributions}

The contributions of this work are the following:

\begin{itemize}
\item We implement the monobit, poker and runs statistical tests in the Python programming language, as well as code that exposes an Arduino to these tests. (Section ~\ref{sec:stattests}.)
\item We provide a program that given a sequence from the \texttt{avr-libc} PRNG seeded with a value from the \texttt{analogRead}-function on an Arduino, determines the seed value. It was done by first analyzing data from the Arduino and building a probability distribution of the values. The program either collects data directly from the Arduino first or can be supplied with a data set. We supply a typical data set with the code. This includes an implementation of the \texttt{avr-libc random} function. (Section ~\ref{sec:seedfind}.)
\item We rebut the claim made by the Arduino manufactures that \texttt{analogRead} returns ``fairly random'' integers ~\cite{ardref}. (Section ~\ref{sec:refuting}.)

\end{itemize}

All of the Ardrand code is free software and is maintained at \url{http://gitorious.org/benediktkr/ardrand}

\section{Related Work -  Background}

Hardware random number generators have been designed with various methods. The search for external entropy has lead researchers down imaginative paths.

Air turbulence in hard drives ~\cite{airturb}, which is proven to be a chaotic phenomenon, has been used as a source for random numbers. The raw disk times where both structured and correlated. The authors used the Fast Fourier transform algorithm to remove bias and correlation. The worst case observed bitrate was 100 bits/minute. 

Intel CPUs contain an on-board RNG ~\cite{intel} chip. It samples thermal noise, shot noise and flicker noise, all of which are present in resistors. The voltage measured across undriven resistors is amplified, but these measurements are correlated to enviromental charasteristics, such as electromagnetic radiation, temperature and power supply fluctuations. The random source used is derived from two free-running oscillators, one fast and one much slower. The thermal noise is used to modulate the frequenciy of the slower clock. The erratic ticks of the slow clock is then used to trigger measurements of the faster one.

Drift between the two clocks is used as a source for binary random digits. The initial measurements are then processed by the von Neumann box. On average, one bit is generated for every 6 raw binary samples. 

The limitations of using hard drives as a source of entropy is that not all computers have hard drives, for instance special-purpose hardware like routers or switches. As solid state drives (SSD) become more available, not even all general purpose computers have spinning hard drives any more. Routers and other special purpose hardware do not have exeternal sources for entropy, other than network traffic, which may be observeable or even controllable by an adversary. 

One example of this is the OpenWRT router ~\cite{lrng}. Since it is based on Linux, it uses the Linux Random Number Generator (LRNG), \texttt{/dev/random} and \texttt{/dev/urandom}. It provides cryptographic services such as SSH, SSL and wireless encryption. It lacks entropy sources other than network interrupts. It has no mouse or keyboard, so it is impossible to use any user interaction to collect entropy. Although this is not part of the LRNG itself, almost all distributions include a script that saves the state of the LRNG between reboots. This is done so that when the operating system starts, the LRNG has a fresh starting state. The OpenWRT distribution does not do this and thus the LRNG state is reset to a predictable state on every reboot, only determined by time of day and a constant string. This example demonstrates that there is need for external entropy sources.

\section{Theoretical Considerations}

We begin by making the notions of ``fairly random'' and ``statistically random'' more precise by defining statistical tests for sequences of integers. Let us first define a few terms, following the exposition by Menezes et al. ~\cite{menezes1996}.

\begin{mydef}
A random bit generator (RBG) is a device or algorithm that outputs a sequence of statistically independent and unbiased binary digits.   
\end{mydef}

A random bit generator can easily be used to generate random numbers.  To obtain an integer in the interval $[0, n]$ we can simply generate $\lfloor \lg n \rfloor + 1$ bits and cast over to an integer. If the result exceeds $n$, one option is to discard it and generate a new number. 

\begin{mydef}
  
  A \emph{pseudorandom random bit generator} (PRBG) is a deterministic algorithm or program that given a truly random binary sequence of length $k$, outputs a binary sequence that appears to be random. The input to the PRGB is called the \emph{seed}, while the output is called a \emph{pseudorandom bit sequence}. 

  % Menezes says l >> k, figure out if he means bitshifting. 
  
\end{mydef}

Note that the output from a PRBG is not random in the colloquial sense of the word. Given the deterministic nature of the algorithm, it will always produce the same sequence for any given seed value. 

\begin{mydef}
Let $s$ be a binary sequence. We say that a \emph{run} in $s$ of length $n$ is a subsequence consisting of either $n$ consecutive 0's or $n$ consecutive 1's. A run is neither preceded or proceeded by the same symbol. We call a run of 1's a block and a run of 0's a gap. 
\end{mydef}

\begin{mydef}
  Let $s$ be a binary sequence of length $n$ such that $s = s_0, \ldots, s_{n-1}$ and let $p_i$ be the probability that $s_i = 1$ for any $i$. Way say that the generator generating $s$ is biased if $p_i \neq \frac 1 2$. 
\end{mydef}

Determining what is random and what is not is a deep philosophical question --- proving mathematically that a generator is indeed generating random bits is impossible ~\cite{menezes1996}. Measuring randomness is as much a philosophical question as it is a mathematical one. There are however statistical tests that allow us to detect certain weaknesses a RBG might have. Note that just because a bit sequence from a generator is accepted by the statistical tests, there is no guarantee that it is indeed random. On the other hand, if it is rejected, we can say with certainty that it is non-random. In other word, when a bit sequence is ``accepted'' it really is ``not rejected''.

\subsection{Statistical significance}

We interpret the results of the statistical tests by means of the $\chi^2$-distributions. It is used in the common $\chi^2$-tests to assess the goodness-of-fit. The $\chi^2$ distribution with $k$ degrees of freedom is given by

\[
f(x, k) =
\begin{cases}
  \frac{1}{2^{k/2}\Gamma(k/2)}\,x^{k/2 - 1} e^{-x/2},  & x \geq 0; \\ 0, & \text{otherwise}.
\end{cases}
\]

where $\Gamma$ is the gamma-function, given by

\[
\Gamma(n) = (n-1)!.
\]

Then we can take our observed data and find an $\chi^2$ statistic, denoted $X^2$, such that

\[
X^2 = \sum_i^k \frac{(O_i - E_i)^2}{E_i}
\]

for all $i$, where $E_i$ denotes the expected number of occurrences and $O_i$ denotes the observed number of occurrences. Then the number $X^2$ tells us about the significance of the test, given a significance level $\alpha$. This is usually done by means of a table of percentiles. 

The degrees of freedom is the number of variables that are free to vary. It is worth noting that if we have $m$ different values in our calculations, we can often figure out the $m^{th}$ variable from the $m-1$ other values, so then we would have $k=m-1$ degrees of freedom. This is often the case for our tests, such as the Monobit test we will define below.

\subsection{Statistical tests}
\label{sec:stattests}

 Here we present a few statistical tests we used. We measured against the specifications set forth in FIPS-140-1 ~\cite{fips140, menezes1996} rather than selecting the significance levels ourselves. The motivation is that the FIPS document effectively sets a standard for the tests to satisfy and we therefore have something to measure against. There are several others tests available and NIST has published paper ~\cite{nist} that outlines a few tests such as the \texttt{DIEHARD}\footnote{See \url{http://stat.fsu.edu/~geo/diehard.html}} test suite, the tests outlined by Donald Knuth in The Art of Computer Programming and the Universal Statistical Test by Mauer ~\cite{mauer}

Let $s = s_0, s_1, \ldots, s_{n-1}$ be a binary sequence of length $n$. A single bitstring of length $n = 20000$ from our generator is subjected to each of these tests. If any one of the tests fail, we conclude that the output of our generator is non-random. 

\subsubsection{Monobit test}

In a random bit sequence, one would expect that the number of 1's and 0's are about the same. This test gives us a statistic on this distribution. Let $n_0$ denote the number of 0's and $n_1$ the number of 1's. We then find the statistic

\begin{equation}
X_1 = \frac{(n_0 - n_1)^2}{2}
\end{equation}

which approximately follows a $\chi^2$ distribution with $1$ degree of freedom (given $n$ and $n_0$ we can easily figure out $n_1$). 

\subsubsection{Poker test}

The poker test tests for certain sequences of five numbers (bits) at a time, similar to a hand in poker. In a random sequence we would expect that each hand would appear approximately the same number of times in $s$. Let $m$ be a positive integer such that 

\[
% \lfloor \frac n m \rfloor \geq 5 \cdot 2^m
\displaystyle \left \lfloor \frac nm \right \rfloor \geq 5 \cdot 2^m
\]

and let $k = \lfloor \frac n m \rfloor$. We divide the sequence $s$ into $k$ disjoint parts of length $m$ and let $n_i$ denote the number of sequences of ``type'' $i$.

For a binary sequence $s_i \in s$, where $|s_i| = m$, we let $n_i$ be the number of sequences where $i$ equals the decimal representation of $s_i$. Note that $0 \leq i \leq 2^m$. 

The statistic used is then

\begin{equation}
X_3 = \frac{2^m}{k} \left( \sum_{i=1}^{2^m} n_i^2 \right) - k
\end{equation}

which approximately follows a $\chi^2$ distribution with $2k - 2$ degrees of freedom (df).
 
\subsubsection{Runs test}

The runs test determines if the number of runs (see \textit{Definition 3}) in $s$ is what is expected of a random sequences. The expected number of gaps, or blocks, of length $i$ in a sequence of length $n$ is

\[
e_i = \frac{n-i+3}{2^{i+2}}.
\]

Let $k$ be equal to the largest integer $i$ for which $e_i \geq 5$, or $k = \max_i e_i \geq 5$. Let $B_i, G_i$ be the number of blocks and gaps, respectively, of length $i$, for each $ 1 \leq i \leq k$. The statistic used is then

\begin{equation}
X_4 = \sum_{i=1}^k \frac{(B_i - e_i)^2}{e_i} + \sum_{i=1}^k \frac{(G_i - e_i)^2}{e_i}
\end{equation}

which approximately follows a $\chi^2$ distribution with 2k-2 degrees of freedom. We note that this exactly finds the $\chi^2$ statistic since the number of runs is the sum of all gaps and blocks. 

\subsection{FIPS140-1 bounds}

We use the FIPS-140-1 bounds ~\cite{fips140} for the tests of our Arduino RBG. Let $s$ be a bit sequence of length 20,000. The documents states explicit bounds as follows:

\begin{description}
\item[Monobit test] The test is passed if $9.654 < X_1 < 10.346$ and the number $n_1$ of 1's should satisfy $9654 < n_1 < 10346$. Should follow a $\chi^2$ with 1 degree of freedom. 
\item[Poker test] The statistic $X_3$ is computed for $m=4$ and the test is passed if $1.03 < X_3 < 57.4$. Should follow a $\chi^2$ with 15 df. 
\item [Runs test] We count the number of blocks and gaps of length $i$ --- $B_i$ and $G_i$ respectively --- in the sequence $s$, for each $1 \leq i \leq 6$. For the purpose of this test, runs longer than 6 are truncated to length 6 ~\cite{fips140}. The test is passed if the number of runs is each within the corresponding intervals below in table ~\ref{tab:fipsbounds}. The bounds must hold for both blocks and gaps, all 12 counts must lie within the bounds. The distribution should follow a $\chi^2$ with 16 df. 

  \begin{table}[H]
    \begin{center}
      \begin{tabular}{| l | l |}
        \hline
        Length of run & Required Interval \\
        \hline
        \hline
        1 & 2267 - 2733 \\
        2 & 1079 - 1421 \\
        3 & 502 - 748 \\
        4 & 223 - 403 \\
        5 & 90 - 223 \\
        6 & 90 - 223 \\
        \hline
      \end{tabular}
    \end{center}
    \caption{Required intervals for runs test as specified by FIPS-140-1}
    \label{tab:fipsbounds}
  \end{table}

\item[Long runs test] The long runs test is passed if there are no runs of length greater than 34 in the bit sequence $s$. 
\end{description}

\subsection{Decorrelation with the von Neumann box}

Decorrelation is a term that refers to reducing autocorrelation, the similarity between observations as a function of the time separation between them. This should not be observed in a random sequence, since the very definition of randomness implies differences in the sequence. A source of randomness may be faulty in that the output of it is either biased or correlated. 

Suppose that the probability that a RBG generates a 1 with a probability $p$ and a 0 with probability $1-p$, where $p$ is unknown but fixed. We group the output of the generator into pairs of two bits. The pairs 00 and 11 are discarded, and a 10-pair is transformed to a 1-bit while a 01-pair is transformed into a 0. This procedure is called the von Neumann-corrector ~\cite{menezes1996, intel} or von Neumann-box. 

\subsection{Algorithms used to try to extract entropy from the Arduino}
\label{sec:algs}

We implemented several algorithms in our search for entropy. These are descriptions of our algorithms. 

The \texttt{Mean-RAND} algorithm is implemented by keeping a list of the $k$ last values and their mean. Then we compare the new reading to the mean and evalute to 0 if it is less, otherwise 1. To remove bias and reduce correlation we run it through the von Neumann-box. 

\begin{lstlisting}[caption=The \texttt{Mean-RAND} algorithm in Python esque pseudocode]
def meanrand(n):
  buf = deque([0]*k)
  for i in [0..k]:
    buf.push(analogRead())

  meanval = sum(buf)/len(buf)

  for i in [0..n]:
    meanval -= buf.pop()/k
    buf.push(analogRead())
    meanval += buf[-1]/k
    m = ceil(meanval)

    yield vNbox(1 if analogRead() > m else 0)
  \end{lstlisting}

The \texttt{Updown-RAND} algorithm first reads an initial value $v_0$ which is then used to determine if the next bit value $v_1$ is 1 if $v_1 > v_0$ and 0 otherwise. We do this twice, i.e. we collect $v_{1,0}$ and $v_{1,1}$ and compare them with the von Neumann box until we obtain a legit bit. This algorithm showed low performance and bandwidth, and has consistently failed the statistical tests. 

\begin{lstlisting}[caption=The \texttt{Updown-RAND} algorithm]
def updownrand(n):
  v0 = analogRead()
  for i in [0..n]:
    yield vNbox(1 if analogRead() > v0 else 0)
\end{lstlisting}

The \texttt{MixMeanUpdown-RAND} algorithm acquires one bit from \texttt{Mean-RAND} and one from \texttt{Updown-RAND} and XORs them together to produce a new bit. Since this method is dependent on \texttt{Updown-RAND} it performs even worse, both in regards to bandwidth and entropy. 

\begin{lstlisting}[caption=The \texttt{MixMeanUpdown-RAND} algorithm]
def mixmeanupdown(n):
  m = meanrand()
  u = updownrand()
  for i in [0..n]:
    yield vNbox(m.next()^u.next())
\end{lstlisting}

Let $a = a_9 \ldots a_1a_0$ be the binary representation of a 10-bit integer read from the \texttt{analogRead}-function on the Arduino. The \texttt{Leastsign-RAND} algorithm simply yields the least significant bit $a_0$. As expected, this algorithm shows greater performance and some promise in regards to randomness. We use the von Neumann-box for decorrelating the output. 

\begin{lstlisting}[caption=The \texttt{Leastsign-RAND} algorithm]
def mixmeanupdown(n):
  for i in [0..n]:
    yield vNbox(analogRead()&1)
\end{lstlisting}

The \texttt{TwoLeastsign-RAND} algorithm works in a very similar fashion. Instead of just using the least significant bit, we use the two least significant bits $a_0$ and $a_1$, XOR them together and run through the von Neumann-box. This algorithm has shown the greatest potential for entropy and has also been implemented on the Arduino itself.

\begin{lstlisting}[caption=The \texttt{TwoLeastsign-RAND} algorithm]
def twoleastsign(n):
  for i in [0..n]:
    yield vNbox(analogRead()&1^(analogRead()>>1)&1)
\end{lstlisting}

\subsection{NIST Security Levels}

National Institute of Standards and Technology (NIST, America) has defined ~\cite{fips140} four basic security levels for cryptographic modules, such as RBGs and RNGs, as well as explicit bounds for statistical tests a RBG must satisfy. The security levels can be outlined as follows

\begin{description}
\item[Security level 1] is the lowest level of security that specifies basic requirements for a cryptographic module. No physical mechanisms are required in the module beyond protection-grade equipment. It allows software cryptography functions to be performed by a regular computer. Examples of systems of level 1 include Integrated Circuit Boards and add-on security products. 

\item[Security level 2] adds the requirement for tamper-proof coatings and seals, or pick-resistant locks. The coatings or seals would be placed on the module so that it would have to be broken in order to attain physical access to the device. It also adds the requirement that a module must authenticate that an operator is authorized to assume as specific role. 

\item[Security level 3] extends the requirements of level 2 to prevent the intruder from gaining access to critical security parameters within the module and if a cover is opened or removed, the critical parameters are erased. 

\item[Security level 4] is the highest level of security. It protects the module from compromise of its security by environmental factors, such as voltage or temperature fluctuations. If one attempts to cut through an enclosing of the module, it should detect this attempt and erase all sensitive data. Most existing products do not meet this level of security. 
\end{description}

Although we were not aiming for physical security in this scenario, aiming for security level 1 seems like a reasonable decision. Note that in order for a device to conform to any of the security modules it has be able to perform self-tests, both at request and start-up. We implemented the tests in the Python programming language on a general-purpose computer. 

FIPS140-1 specifies that the sample must be 20,000 bits, or 2.5KB. But the Arduino Duemilanove only has 2 KB of RAM. Luckily, it has a 32KB Flash memory which could be utilized to implement the statistical tests on the Arduino itself.

\section{Experimental Results}

\begin{figure}[h!]
    \centering  
    \subfloat[Sample from Ard3 taken on the desktop computer]{
      \label{fig:allpinsdesktop}
      \includegraphics[width=0.55\columnwidth]{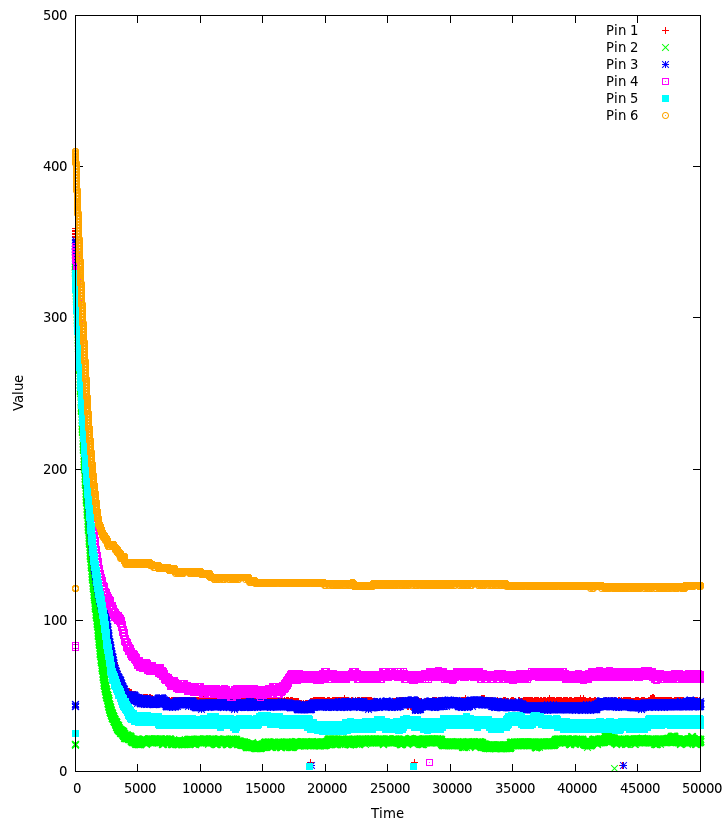}
    }
    \subfloat[Sample from Ard3 taken on the D620]{
      \label{fig:allpinsd620}
      \includegraphics[width=0.55\columnwidth]{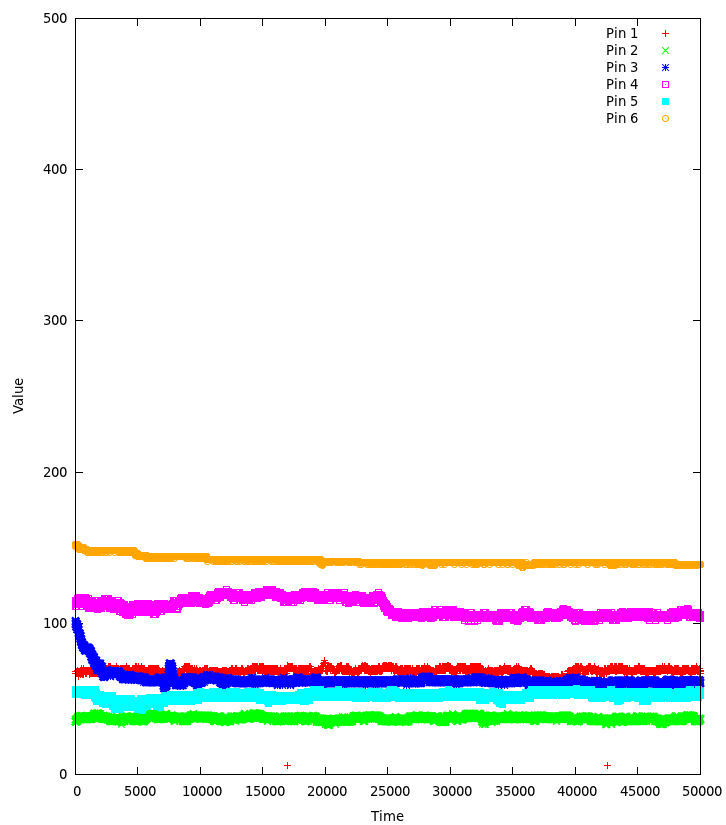}
    }

    \caption{Readings from Ard3 taken over all pins on both a desktop computer and a D620 laptop.}
    \label{fig:allpins}
        
  \end{figure}

We began by analyzing the output of the function \texttt{analogRead} on the Arduino in various different settings. We found that the output is dependent on several environmental factors, some of which are unknown to us. Reading from different pins gives different scopes of values, but the behavior is the same. We will show graphs of all pins on one Arduino connected to two computers to back this claim, see Figure ~\ref{fig:allpins}. The computer to which the Arduino is connected to affects the results. On one computer tested, one of our algorithms produces sequences that were not rejected by the statistical tests. 

We want to know if, and by how much, the environment affects the results. We also investigate how and if we can use the \texttt{analogRead} output to find entropy and how we test for randomness. 

\subsection{Computers and devices used in research}

The output of the function \texttt{analogRead} on the Arduino is somewhat dependent on the environment in which it resides. We subjected the Arduino boards to different conditions, such as putting it in the freezer or on top of a hot heating element. 

These are descriptions of the computers used for the experiments. 

% \begin{description}
% \item[Setting 0] The study of the author. An apartment built by the U.S. Navy with electricity converted to Icelandic and European standards. 
% \item[Setting 1] Garage furnished as a studio apartment in Kópavogur. 
% \item[Setting 2] Computer Science lab at Reykjavik University. 
% \end{description}

\begin{itemize}
\item A no-name desktop computer with a Gigabyte GA-MA69GM-S2H motherboard. This machine runs the Debian GNU/Linux testing/wheezer operating system.
\item A Dell D505 laptop. This machine runs the Ubuntu GNU/Linux Maverick Meekat 11.04 operating system.
\item A Dell D620 laptop. This machine runs the Ubuntu GNU/Linux Maverick Meekat 11.04 operating system as well.
\end{itemize}

We found that both the laptops showed same or similar behavior, both on raw outputs and statistical testing, but the desktop differed. It is unclear what aspects trigger the deviations but we will discuss this point further in section ~\ref{sec:harvest}. For our research we used three identical Arduino Duemilanove boards with the ATMega 328 micro-controller. To distinguish between them, we call them ard1, ard2 and ard3. These names are also used to distinguish between them in our data samples. 

\subsection{Analysis of \texttt{analogRead}}

Our first hypothesis was the space and volume of the area that the Arduino resided in affected the values. If we look at Figure ~\ref{fig:ard3space} we can see that the environment lays some role and that where you place it definitely has effect on the output. The figure shows readings taken in various different places --- in an open space, a closed cupboard, inside a computer case and in a larger open space. 

\begin{figure}[H]
  \centering  
  \includegraphics[width=0.7\columnwidth]{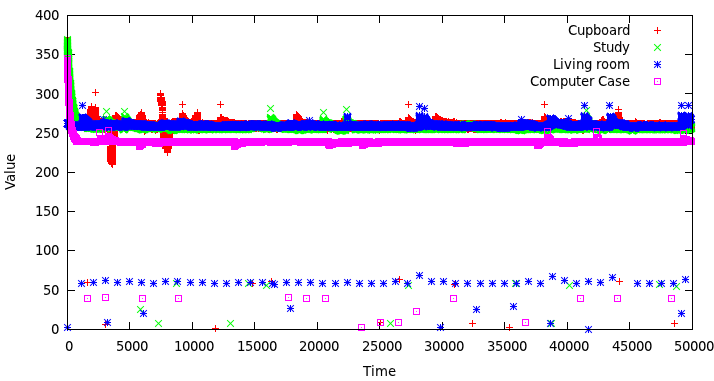}
  
  \caption{Readings from ard1 connected to the desktop. Samples are taken inside a small cupboard, in a fairly large room, a large living room and inside the desktop computer case itself.}
  \label{fig:ard3space}
\end{figure}

\begin{figure}[h!]
  \centering
  \includegraphics[width=0.7\columnwidth]{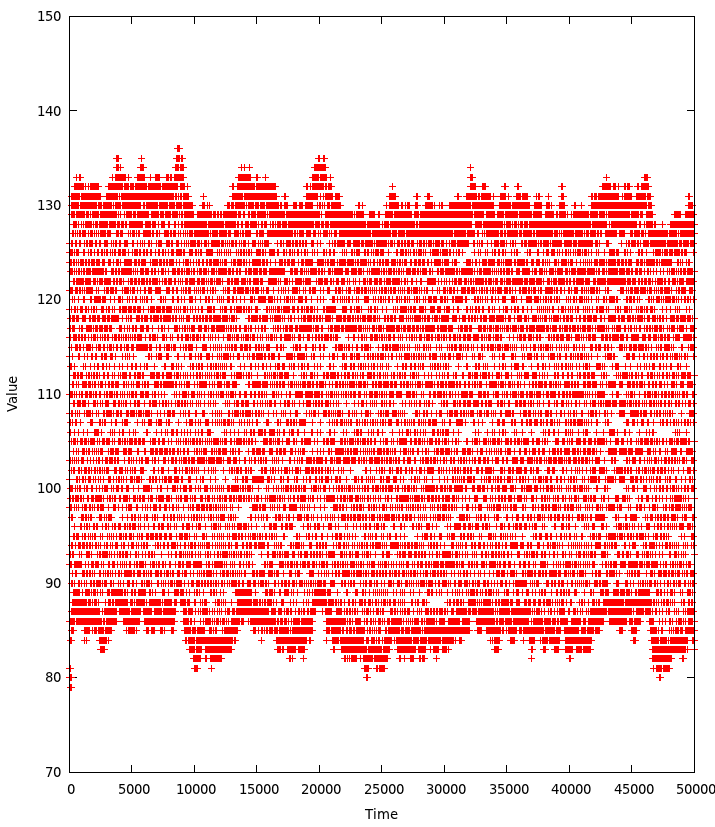}
  \caption{Readings from ard3 connected to the D620 laptop}
  \label{fig:gfgarageard3}
\end{figure}

Moreover, by looking at the graph it becomes evident that there is limited entropy available. Note the drop at the beginning; it does not appear for all computers, e.g. specifically when connected to the D620 laptop (see Figure ~\ref{fig:gfgarageard3}). It should be noted that the data originates from the same Arduino device, in the same setting. The only factor is the computer used.

\begin{figure}[h!]
  % \centering  
  % \subfloat[200 readings from ard1 connected to the desktop computer]{
  %   \label{fig:ard1zoomroom}
  %   \includegraphics[width=0.7\columnwidth]{Room_1500-1700_zoom.png}
  % }

  % \subfloat[1000 readings from ard3 connected to the D620 laptop]{
  %   \label{fig:ard3zoomroom}
  %   \includegraphics[width=0.7\columnwidth]{RoomZoom_9des.png}
  % }

  % \caption{Limited ranges of readings from \texttt{analogRead}}
  % \label{fig:ardzoom}
  \centering
  \includegraphics[width=0.7\columnwidth]{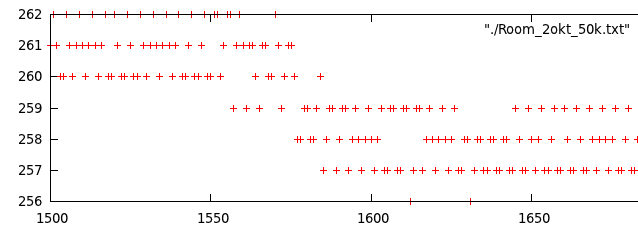}
  \caption{200 readings from ard1 connected to the desktop computer}
  \label{fig:ardzoom}

\end{figure}

Our experiments have shown that the output is fairly regular and if we look at Figure ~\ref{fig:ardzoom}, showing more limited ranges of readings, we see that the structure and apparent lack of entropy. The readings should have been heavily influenced by analog noise ~\cite{ardref}. This is further investigated in section ~\ref{sec:refuting}. 

Note the interference patterns in e.g. figures ~\ref{fig:ard1freezer} and \ref{fig:ard1heat} --- they show up more clearly in the case of the temperature experiments since we see a much wider range of values. Although we are not sure what causes these patterns, electrical fluctuations are a potential candidate. These patterns might also be a product of the analog pins themselves, or their manufacturing process. The exact physical causes for this phenomenon appear complex and are beyond the scope of this paper. 

\FloatBarrier
\subsubsection{Effects of temperature}
\FloatBarrier

\begin{figure}[h!]
  \centering  
  \includegraphics[width=0.6\columnwidth]{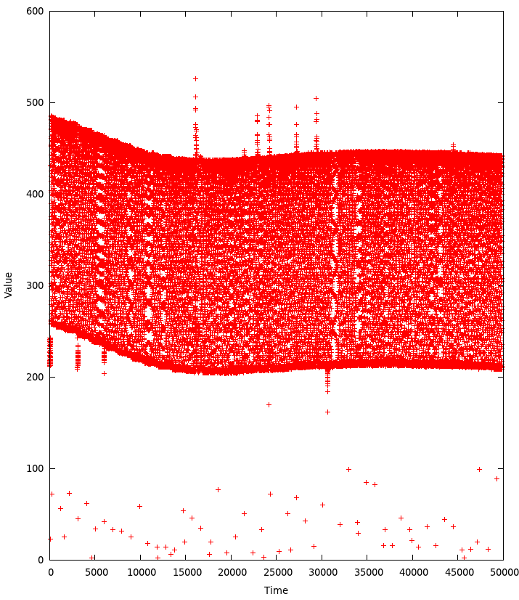}
  \caption{Readings on top of a hot heating element (approx. 40C) connected to a Dell D505}
  \label{fig:ard1heat}
\end{figure}

\begin{figure}[h!]
    \centering  
    \subfloat[Fridge (approx. 1C)]{
      \includegraphics[width=0.5\columnwidth]{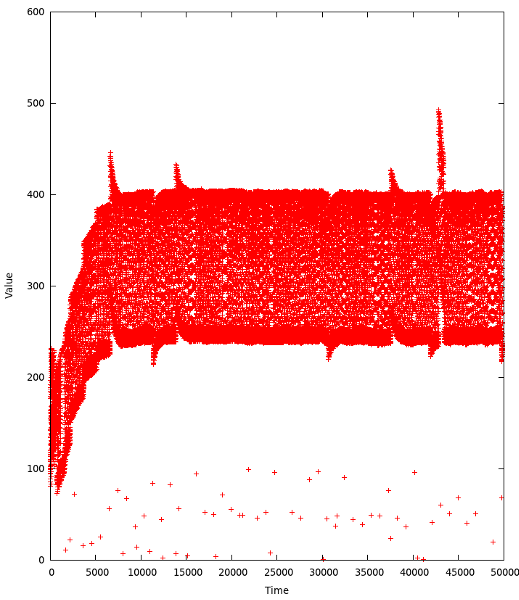}
      \label{fig:ard1fridge}
    }
    \subfloat[Freexer (approx. -11C)]{
      \includegraphics[width=0.5\columnwidth]{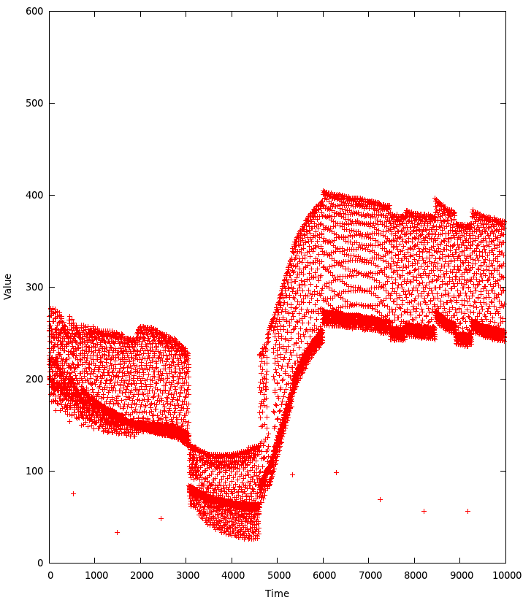}
      \label{fig:ard1freezer}
    }

    \caption{Readings in cold temperatures with a Dell D505 laptop}
    \label{fig:ardcold}
        
  \end{figure}

Temperature is a key environmental factor. We see a much broader range of values when the Arduino is operating in heat or cold. The figures ~\ref{fig:ard1heat}, \ref{fig:ard1freezer} and ~\ref{fig:ard1fridge} show the output from \texttt{analogRead} in various temperature conditions of the extreme kinds. Note that Figure ~\ref{fig:ard1freezer} only has 10000 values, as opposed to the 50000 values in all the other figures. This is because the Arduino simply stops working after a few minutes at $-11^\circ$C. Arduino have not released any information regarding operating temperatures but according to AVR the operating temperatures for the ATMega 328 micro-controller is $-40^\circ$C to $85^\circ$C  ~\cite{atmegads}. One of our Arduino boards (ard1) broke after spending 4 hours in the freezer at $-12^\circ$C, so we conclude that some other component(s) on the board survive less cold than the micro-controller itself.

%\begin{figure}[h!]
%  \centering  
%  \includegraphics[width=0.6\columnwidth]{freezer.png}
%  \caption{Reading inside a freezer (approx. -11C) at setting connected to a Dell D620w}
%  \label{fig:ard1freezer}
%\end{figure}
  
%\begin{figure}[h!]
%  \centering  
%  \includegraphics[width=0.6\columnwidth]{fridge.png}
%  \caption{Reading inside a fridge (approx. 1C) at setting connected to a Dell D505}
%  \label{fig:ard1fridge}
%\end{figure}

\FloatBarrier
\subsection{Harvesting entropy and statistical testing}
\label{sec:harvest}

We subject the output of the \texttt{RAND} algorithms described in section ~\ref{sec:algs} to the statistical tests described in section ~\ref{sec:stattests}. These are our results using the three different computers used in our experiments. 

We used a baudrate of 115200 bps, fast available for the Arduino over the FTDI RS232-to-USB connection, in these experiments for maximum bandwidth. The maximum reading rate of \texttt{analogRead} is about 10000 times per second ~\cite{ardref}. When we read\footnote{See \texttt{poll.py} in the Ardrand codebase} iteratively from \texttt{analogRead} over USB,  \texttt{pyserial} will raise either an \texttt{OSError} or \texttt{SerialException}. This happens approximately once every 500 times and reading the \texttt{pyserial} source code\footnote{See \url{http://www.java2s.com/Open-Source/Python/Development/pySerial/pyserial-2.5-rc2/serial/serialposix.py.htm}} we find a comment stating that ``disconnected devices, at least on Linux, show the behavior that they are always ready to read immediately but reading returns nothing''. We therefore conclude that these exceptions are the result of \texttt{analogRead} blocking for moment. 

These are the results of our \texttt{RAND} algorithms using all three computers, subjected to the FIPS boundaries. Each bit sequence tested is $n = 20000$ bits long and each recorded result is the average of three consecutive runs. Green means Accepted and red Rejected. 

\subsubsection{Results with Desktop computer}

  \begin{table}[H]
    \begin{center}
      \begin{tabular}{| l | p{1.8cm} | p{2.1cm} | l | l | l |}
        \hline
        Algorithm  & Monobit & Poker & Runs & Long runs & Bandwidth\\
        \hline
        \hline
        \texttt{Leastsign} & \textcolor{green}{$n_1 = 9947$}
        & \textcolor{red}{$X_3 = 869.44$}
        & \textcolor{red}{\textbf{Rejected}}
        & \textcolor{green}{\textbf{Accepted}}
        & 290.55 bps \\

        \texttt{Twoleastsign} & \textcolor{green}{$n_1 = 10027$}
        & \textcolor{red}{$X_3 = 1290.05$}
        & \textcolor{red}{\textbf{Rejected}}
        & \textcolor{green}{\textbf{Accepted}}
        & 133.6 bps \\

        \texttt{Mean} & \textcolor{green}{$n_1 = 9979$}
        & \textcolor{red}{$X_3 = 149.87$}
        & \textcolor{red}{\textbf{Rejected}}
        & \textcolor{green}{\textbf{Accepted}}
        & 85.34 bps \\

        \texttt{Updown} & \textcolor{red}{$n_1 = 8352$}
        & \textcolor{red}{$X_3 = 1959.2$}
        & \textcolor{red}{\textbf{Rejected}}
        & \textcolor{green}{\textbf{Accepted}}
        & 3.87 bps \\

        \hline
      \end{tabular}
    \end{center}
    \label{tab:res0}
    \caption{Statistical tests on desktop}
  \end{table}

\subsubsection{Results with the D620 laptop}

  \begin{table}[H]
    \begin{center}
      \begin{tabular}{| l | p{1.8cm} | p{2.2cm} | l | l | l |}
        \hline
        Algorithm & Monobit & Poker & Runs & Long runs & Bandwidth\\
        \hline
        \hline
        \texttt{Leastsign} & \textcolor{green}{$n_1 = 10 006$}
        & \textcolor{green}{$X_3 = 34.59$}
        & \textcolor{red}{\textbf{(Rejected)}}
        & \textcolor{green}{\textbf{Accepted}}
        & 290.55 bps \\

        \texttt{Twoleastsign} & \textcolor{green}{$n_1 = 10 027$}
        & \textcolor{green}{$X_3 = 10.36$}
        & \textcolor{green}{\textbf{Accepted}}
        & \textcolor{green}{\textbf{Accepted}}
        & 172.0 bps \\

        \texttt{Mean} & \textcolor{green}{$n_1 = 10030$}
        & \textcolor{red}{$X_3 = 4743,17$}
        & \textcolor{red}{\textbf{Rejected}}
        & \textcolor{green}{\textbf{Accepted}}
        & 25.32 bps \\

        \hline
      \end{tabular}
    \end{center}
    \label{tab:res1}
    \caption{Statistical test on D620 laptop}
  \end{table}

We can see that the \texttt{Twoleastsign-RAND} algorithm here produces sequences that are not rejected as being non-random. The \texttt{Leastsign-RAND} algorithm is a little less consistent since it is rejected by the tuns tests some of the time.

We have exposed sequences that pass our statistical tests to a statistical test suite made available by NIST\footnote{See \url{http://csrc.nist.gov/groups/ST/toolkit/rng/index.html}}. This suite consists of 15 tests, some of which are also implemented by us. We found that when a sequences passes our tests, it will also pass all the NIST tests. This implies that it is possible to generate random bits on the Arduino, but it relies on external factors that are not fully understood by us. Our best guess is that the voltage from the USB connection on the computer influences the regularity of the \texttt{analogRead} readings. 

 As we can see in Figure ~\ref{fig:gfgarageard3} and ~\ref{fig:allpinsd620}, \texttt{analogRead} is producing different patterns (and no initial drop) on this computer, compared to the desktop computer. Note the drop on pin 3 in the beginning. Curiously, the \texttt{Twoleastsign-RAND} algorithm will fail when we choose pin 3. There is no guarantee that these sequences are truly random, they have just not been rejected as non-random. This shows that using some computers, the Arduino could possibly work as a RBG. But for a device to be a RBG, it has to work using all PC hardware, in all settings.

\subsubsection{Results with the D505 laptop}

 \begin{table}[H]
    \begin{center}
      \begin{tabular}{| l | p{1.8cm} | p{2.2cm} | l | l | l |}
        \hline
        Algorithm & Monobit & Poker & Runs & Long runs & Bandwidth\\
        \hline
        \hline
        \texttt{Leastsign} & \textcolor{green}{$n_1 = 10 033$}
        & \textcolor{green}{$X_3 = 27.4$}
        & \textcolor{red}{\textbf{Failed}}
        & \textcolor{green}{\textbf{Accepted}}
        & 473.32 bps \\

        \texttt{Twoleastsign} & \textcolor{green}{$n_1 = 10080$}
        & \textcolor{green}{$X_3 = 18.68$}
        & \textcolor{green}{\textbf{(Accepted)}}
        & \textcolor{green}{\textbf{Accepted}}
        & 240.0 bps \\

        \texttt{Mean} & \textcolor{green}{$n_1 = 9980$}
        & \textcolor{red}{$X_3 = 3365.65$}
        & \textcolor{red}{\textbf{Rejected}}
        & \textcolor{green}{\textbf{Accepted}}
        & 27.1 bps \\

        \hline
      \end{tabular}
    \end{center}
    \label{tab:res1}
    \caption{Statistical with the D505 laptop}
  \end{table}

This computer has shown inconsistent test results and sequences produced by it will sometimes be accepted by the poker and runs tests, while sometimes they are not. As before, the causes for this are unknown to us and open for speculations.

\section{Breaking the Arduino as a RNG}

This section is twofold. We will both show that using the \texttt{analogRead} function to seed the \texttt{avr-libc} PRNG does not give adequate security and we also exhibit proof-of-concept code that finds such a seed value, given a sequence from the PRNG. 

\subsection{Refuting the claims made by Arduino}
\label{sec:refuting}

The Arduino Reference Manual ~\cite{ardref} states the following in the section about the \texttt{randomSeed} function. This claim is at the time of writing found in the manual, and is available via The Internet Archive\footnote{See \url{http://web.archive.org/web/20110428064453/http://arduino.cc/en/Reference/RandomSeed}}. The reference manual is only available online. 

\begin{quote}
``If it is important for a sequence of values generated by random() to differ, on subsequent executions of a sketch, use randomSeed() to initialize the random number generator with a fairly random input, such as analogRead() on an unconnected pin.''
\end{quote}

After having visually examined the raw output with the graphs in the section above, we clearly saw that the output is very likely non-random and not even ``fairly random'' as claimed. This would also explain the troubles we had in devising an algorithm that produces random bits. 

The first issue with \texttt{analogRead} is that it only returns 10-bit integers, since it reads from the 10-bit analog-to-digital converter on the Arduino board\footnote{See \url{http://arduino.cc/en/Reference/AnalogRead}}. It then follows trivially that if you use \texttt{analogRead} to seed the PRNG, there are only $2^{10} = 1024$ seed values for an adversary to explore. 

As we can see from Figure~\ref{fig:allpins} there are only roughly 100-400 values that show up, but it is worth noting that using different pins give us different scopes of values. Note the drop when using the desktop but not the D620 laptop. 

We exposed the output from \texttt{analogRead} to the same statistical tests as our \texttt{RAND} algorithms. In order to use the FIPS-bounds to measure against we needed 20 000 bits, or 2000 10-bit integers that we converted to binary. We state the null hypothesis as follows, 

\[
H_0 = \text{The output from \texttt{analogRead} is "statistically random"}
\]

and show that the results are statistically significant and we can reject it as non-random. Note that the Arduino developers claim that the output is ``fairly random'' and not ``statistically random''. These are the average over three consecutive runs in setting 0. 

  \begin{table}[H]
    \begin{center}
      \begin{tabular}{| l | l | l | l |}
        \hline
        Statistical test & $X$ statistic & Accepted & Required $X$ interval \\
        \hline
        \hline
        \textbf{Monobit} & 903.847 & \textcolor{red}{\textbf{Rejected}} & $9.964 < X_1 < 10.346$  \\
        \textbf{Poker} & 3211.45 & \textcolor{red}{\textbf{Rejected}} & $1.03 < X_3 < 57.4$ \\
        \textbf{Runs} & 2812.81 & \textcolor{red}{\textbf{Rejected}} & Lengths of runs used \\
        \hline
      \end{tabular}
    \end{center}
    \caption{Results of the statistical test applied to \texttt{analogRead} output}
    \label{tab:analogreadtests}
  \end{table}

We see that all of the statistics are far off from the FIPS requirements so we can safely conclude that the null hypothesis $H_0$ is false and \texttt{analogRead} is not even ``fairly random''. 

We note that the observed bitrate for reading values directly from \texttt{analogRead} is 17531 bps\footnote{Baudrate 115200 bps}. 

\subsection{Finding the seed}
\label{sec:seedfind}

This limited range of possible values from \texttt{analogRead} cuts down on search time for the seed. As we have seen, there are only a few hundred values that show up most of the time, although these values may vary. We have designed proof-of-concept code\footnote{Implemented in the file \texttt{seedfind.py} in the Ardrand codebase} that given a sequence from the \texttt{avr-libc} (Arduino) PRNG, finds the seed --- assuming it was generated by \texttt{analogRead}. 

The \texttt{avr-libc} PRNG is a Linear congruential generator defined by the recurrence relation\footnote{Resides in \texttt{libc/stdlib/random.c} in avr-libc-1.7.1 and a Python implementation is found at \texttt{avrlibcrandom.py} in the Ardrand codebase}

\[
X_{n+1} \equiv 7^5 \cdot X_n \pmod{2^{31} -1}.
\]

In order to account for the diversity in values that the Arduino returns in various settings, our implementation inputs either a text file of samples or can connect to an Arduino board and collect fresh samples. To provide the same interface, this is implemented by means of inherited classes in the code. 

Let $C$ denote the number of calls a program has made to the PRNG in order to generate a sequence $s$. Let $m$ be our best-guess or estimation of the unknown $C$. Our program inputs $s$ and $m$, as well as a sample source. It then builds a list of values in the range $[0, 1023]$, sorted by the frequency by which they appear in the given sample. Let $P$ denote this list of probability distribution values.

We then create one bidirectional queue for each of the $1024$ values in P. Let $k$ be the length of the sequence $s$. For all $i \in P$ we create a deque with $k$ pseudo random integers derived from $i$ as a seed. Then we iterate over $P$ and generate $k+m$ integers for each deque (holding $k$ values at a time) until we find the sequence. 

Here is pseudo-code for our program. The functions \texttt{srandom} and \texttt{random} are the seeding function for the \texttt{avr-libc} PRNG and random function, respectively. 

\begin{lstlisting}[caption=Finding the seed]
def findseed(s, m, samplesource):
  k = len(s)
  lastk = [deque()]*1024

  P = buildProbDist(samplesource)

  # Expand all the deques by k elements from P[i] as seed
  for i in P:
    srandom(i)
    lastk[i] = deque([random() for _ in range(k)])
    # Did we receive a sequence derived directly from the seed?
    if lastk[i] == s:
      return i

  while True:
    for i in P:
      for _ in range(m+k):
        srandom(lastk[i][-1])
        v = random()
        lastk[i].popfront()
        lastk[i].append(v)
        if lastk[i] == s:
          return i
\end{lstlisting}

This program has running bounds given by $\mathcal{O}(C)$, since it has a endless \texttt{while}-loop, only bounded by $C$. The running time is thus not bounded by the $\mathcal{O}(m+k)$ loop, since that is only an estimation or best-guess of how long it takes to find the correct seed. 

\subsubsection{Possible optimizations}

While this program runs reasonably fast in practice, one can think of optimizations of the code. Let $G$ a sorted list of observed values in the sample source. Then one variation of the \texttt{findseed} program might generate sequences from each value $g \in G$ as seed for some constant time $t$ before moving on to the less likelier values that are in $P-G$, the unobserved values. Thus we would spend $t$ times more time on the more probable values. The while loop of this variation would look like

\begin{lstlisting}[caption=One possible optimization of the \texttt{findseed} program]
...
  while True:
    # Iterate over a sorted list of the observed values
    for i in G:
      for _ in range(t*(m+k)):
        expand(lastk[i])
        if lastk[i] == s:
          return i
    # Check the unobserved values, but spend less time there
    for i in P-G:
      for _ i range(m+k):
        expand(lastk[i])
         if lastk[i] == s:
           return i

def expand(que):
  srandom(que[-1])
  v = random()
  que.popfront()
  lastk[i].append(v)
\end{lstlisting}

In order to determine the efficiency of our program, we first pseudo-randomly select an integer $d$ such that $1 \leq d \leq 1000$ with the Python PRNG. Then we generate a sequence $s=s_1, \ldots, s_{d+100}$ with the \texttt{avr-libc} and pass the subsequence $s_{d+1}, \ldots, s_{d+100}$ to our program. To seed the PRNG, we use the Python PRNG to select a \texttt{analogRead} value from a file of samples collected from an Arduino. We did this 5000 times and our program found the seed in every case, with a mean time of 1.6 seconds\footnote{Using a computer with an AMD Athlon 64 X2 4000+ Dual Core Processor} spent on each sequence. 

\section{Conclusions}

The primary goal of this project was to investigate the feasibility of using vanilla Arduino Duemilanove boards without add-on hardware as cheap RBG devices, in hopes of building a true hardware random number generator. Early on, we realized that there simply wasn't enough entropy available to build a RBG. Most to our surprise, we found that using some of our computers, the Arduino seems to work as a RBG when using the \texttt{Twoleastsign-RAND} algorithm. The resulting strings were not rejected by our implemented statistical tests that had rejected all other sequences we tried. It should be noted that due to the very nature of randomness, we cannot say for sure that it is indeed producing random bits --- we can only state that it was not rejected as non-random. But since this is only the case using specific hardware, we must reject the Arduino as a RBG since it is not universal.

Since it is only not rejected using a certain hardware, we have ultimately shown that these boards are not ideal devices for this task. Instead we have refuted the claim made by Arduino themselves and devised a program to find the seed when the onboard PRNG is seeded with \texttt{analogRead}, as recommended by Arduino. 

In future work, the exact nature of the ports read by\texttt{analogRead} should be investigated with respect to the environment or USB connectors. It also remains open to find a cheap, readily available hardware that produce random statistically unbiased random numbers quickly.

\bibliographystyle{plain}
\bibliography{ardrand}

\end{document}